\documentstyle[sprocl]{article}

\def\be{\begin{equation}}
\def\ee{\end{equation}}
\def\bea{\begin{eqnarray}}
\def\eea{\end{eqnarray}}

\begin{document}

\title{BACK REACTION OF COSMOLOGICAL PERTURBATIONS \footnote{BROWN-HET-1212, Jan. 2000; invited lecture at COSMO-99 (ICTP, Trieste, Sept. 27 - Oct. 2, 1999), to appear in the proceedings (World Scientific, Singapore, 2000).}}

\author{ROBERT H. BRANDENBERGER}

\address{Department of Physics, Brown University,\\ 
Providence, RI 02912, USA\\E-mail: rhb@het.brown.edu}


\maketitle\abstracts{The presence of cosmological perturbations affects the background metric and matter configuration in which the perturbations propagate. This effect, studied a long time ago for gravitational waves, also is operational for scalar gravitational fluctuations, inhomogeneities which are believed to be more important in inflationary cosmology. The back-reaction of fluctuations can be described by an effective energy-momentum tensor. The issue of coordinate invariance makes the analysis more complicated for scalar fluctuations than for gravitational waves. We show that the back-reaction of fluctuations can be described in a diffeomorphism-invariant way. In an inflationary cosmology, the back-reaction is dominated by infrared modes. We show that these modes give a contribution to the effective energy-momentum tensor of the form of a negative cosmological constant whose absolute value grows in time. We speculate that this may lead to a self-regulating dynamical relaxation mechanism for the cosmological constant. This scenario would naturally lead to a finite remnant cosmological constant with a magnitude corresponding to $\Omega_{\Lambda} \sim 1$.}

\section{Introduction}

It is well known that gravitational waves
propagating in some background space-time affect the dynamics of
the background. This back-reaction can be described in terms of an effective energy-momentum tensor $\tau_{\mu \nu}$. In the short wave limit, when the typical wavelength of the waves is small compared with the curvature of the background space-time, $\tau_{\mu \nu}$ has the form of a radiative fluid with an equation of state $p= \rho / 3$ (where $p$ and $\rho$ denote pressure and energy density, respectively).  

In most models of the early Universe, scalar-type metric perturbations are
more important than gravity waves. Here, we report on a study of the back reaction problem for both scalar and tensor gravitational perturbations \cite{MAB96,ABM97}. We derive the effective energy-momentum tensor $\tau_{\mu \nu}$ which describes the back-reaction and apply the results to inflationary cosmology.

A crucial issue to be addressed when studying scalar-type metric perturbations is the problem of gauge artifacts. As is well known (see e.g. Mukhanov et al. \cite{MFB92} for a comprehensive review), cosmological perturbations transform non-trivially
under coordinate transformations (gauge transformations). However, the
answer to the question ``how important are perturbations for the evolution
of a background'' must be independent of the choice of gauge, and hence the
back-reaction problem must be formulated in a gauge invariant way.
In  recent work \cite{MAB96,ABM97}, we demonstrated that the back reaction
problem can be set up in a way which is gauge-invariant (under linear coordinate transformations). This summary will not dwell on the issue of gauge-invariance, and the reader is referred to the original articles for details.

Here, we will briefly summarize the formalism and derive the effective energy-momentum tensor $\tau_{\mu \nu}$ which describes the back-reaction. We apply the results to a simple model of chaotic inflation and  discuss the equation of state corresponding to the resulting $\tau_{\mu \nu}$. As we shall see, long wavelength fluctuations dominate, and in the limit where the short wavelength fluctuations can be neglected, the resulting equation of state of $\tau_{\mu \nu}$ is $p\approx -\rho $ with $\rho < 0$, i.e. the equation of state corresponding to a negative cosmological constant. We speculate on the possible connection of this result with a dynamical relaxation mechanism for the cosmological constant (see also Tsamis \& Woodard \cite{TW} for related work).

\section{Gravitational Back-Reaction}

The analysis of gravitational back-reaction \cite{MAB96}  is related to early work by Brill, Hartle and Isaacson \cite{Brill}, among others. The idea is to expand the Einstein equations to second order in the perturbations, to assume that the first order terms satisfy the equations of motion for linearized cosmological perturbations \cite{MFB92} (hence these terms cancel), to take the spatial average of the remaining terms, and to regard the resulting equations as equations for a new homogeneous metric $g_{\mu \nu}^{(0, br)}$ which includes the effect of the perturbations to quadratic order:
\be \label{breq}
G_{\mu \nu}(g_{\alpha \beta}^{(0, br)}) \, = \, 8 \pi G \left[ T_{\mu \nu}^{(0)} + \tau_{\mu \nu} \right]\,
\ee
where the effective energy-momentum tensor $\tau_{\mu \nu}$ of gravitational back-reaction contains the terms resulting from spatial averaging of the second order metric and matter perturbations:
\be \label{efftmunu}
\tau_{\mu \nu} \, = \, < T_{\mu \nu}^{(2)} - {1 \over {8 \pi G}} G_{\mu \nu}^{(2)} > \, ,
\ee
where pointed brackets stand for spatial averaging, and the superscripts indicate the order in perturbations.

As formulated in (\ref{breq}) and (\ref{efftmunu}), the back-reaction problem is not independent of the coordinatization of space-time and hence is not well defined. It is possible to take a homogeneous and isotropic space-time, choose different coordinates, and obtain a non-vanishing $\tau_{\mu \nu}$. This ``gauge" problem is related to the fact that in the above prescription, the hypersurface over which the average is taken depends on the choice of coordinates. 

The key to resolving the gauge problem \cite{MAB96} is to realize that to second order in perturbations, the background variables change. This change can be calculated consistently, and given this change of background quantities it can be shown that the back-reaction problem is formulated in a covariant way by (\ref{breq}) and (\ref{efftmunu}). A gauge independent form of the back-reaction equation (\ref{breq}) can  be derived  by defining background and perturbation variables $Q = Q^{(0)} + \delta Q$ which do not change under linear coordinate transformations. Here, $Q$ represents collectively both metric and matter variables. The gauge-invariant perturbation quantities are Bardeen's gauge-invariant variables \cite{Bardeen}. The gauge-invariant form of the back-reaction equation then looks formally identical to (\ref{breq}), except that all variables are replaced by the corresponding gauge-invariant ones. We will follow the notation of \cite{MFB92}, and use as gauge-invariant perturbation variables the Bardeen potentials \cite{Bardeen} $\Phi$ and $\Psi$ which in longitudinal gauge coincide with the actual metric perturbations $\delta g_{\mu \nu}$. Calculations hence simplify greatly if we work directly in  longitudinal gauge. Recently, the back-reaction analysis of \cite{MAB96,ABM97} has been confirmed \cite{WA} by working in a completely different gauge, making use of the covariant quantization approach.

For simplicity, we shall take matter to be described in terms of a single scalar field. In this case, there is only one independent metric perturbation variable, and in longitudinal gauge the perturbed metric can be written in the form
\be \label{metric}
ds^2 =  (1+ 2 \phi) dt^2 - a(t)^2(1 - 2\phi) \delta_{i j} dx^i dx^j  \, ,
\ee
where $a(t)$ is the cosmological scale factor. The energy-momentum tensor for a scalar field is 
\be 
T_{\mu \nu }=\varphi _{,\mu }\varphi _{,\nu }-g_{\mu \nu }\left[ {\frac 12}%
\varphi ^{,\alpha }\varphi _{,\alpha }-V(\varphi )\right] \,.
\ee

By expanding the Einstein tensor and the above energy-momentum tensor to second order in the metric and matter fluctuations $\phi$ and $\delta \varphi$, respectively, it can be shown that the non-vanishing components of the effective back-reaction energy-momentum tensor $\tau_{\mu \nu}$ become
\bea  \label{tzero}
\tau_{0 0} &=& \frac{1}{8 \pi G} \left[ + 12 H \langle \phi \dot{\phi} \rangle
- 3 \langle (\dot{\phi})^2 \rangle + 9 a^{-2} \langle (\nabla \phi)^2
\rangle \right]  \nonumber \\
&+& {1 \over 2} \langle ({\delta\dot{\varphi}})^2 \rangle + {1 \over 2} a^{-2} \langle
(\nabla\delta\varphi)^2 \rangle  \nonumber \\
&+& {1 \over 2} V''(\varphi_0) \langle \delta\varphi^2 \rangle + 2
V'(\varphi_0) \langle \phi \delta\varphi \rangle \quad ,
\eea
and 
\bea  \label{tij}
\tau_{i j} &=& a^2 \delta_{ij} \left\{ \frac{1}{8 \pi G} \left[ (24 H^2 + 16 
\dot{H}) \langle \phi^2 \rangle + 24 H \langle \dot{\phi}\phi \rangle
\right. \right.  \nonumber \\
&+& \left. \langle (\dot{\phi})^2 \rangle + 4 \langle \phi\ddot{\phi}\rangle
- \frac{4}{3} a^{-2}\langle (\nabla\phi)^2 \rangle \right] + 4 \dot{{%
\varphi_0}}^2 \langle \phi^2 \rangle  \nonumber \\
&+& {1 \over 2} \langle ({\delta\dot{\varphi}})^2 \rangle - {1 \over 6} a^{-2} \langle
(\nabla\delta\varphi)^2 \rangle - 
4 \dot{\varphi_0} \langle \delta \dot{\varphi}\phi \rangle  \nonumber \\
&-& \left. {1 \over 2} \, V''(\varphi_0) \langle \delta\varphi^2
\rangle + 2 V'( \varphi_0 ) \langle \phi \delta\varphi \rangle
\right\} \quad ,
\eea
where $H$ is the Hubble expansion rate.

\section{Equation of State of Back-Reaction in Inflationary Cosmology}

The metric and matter fluctuation variables $\phi$ and $\delta \varphi$ are linked via the Einstein constraint equations, and hence all terms in the above formulas for the components of $\tau_{\mu \nu}$ can be expressed in terms of two point functions of $\phi$ and its derivatives. The two point functions, in turn, are obtained by integrating over all of the Fourier modes of $\phi$, e.g.
\be \label{tpf}
\langle \phi^2 \rangle \, \sim \,  \int_{k_i}^{k_u} {dk} k^2 \vert \phi_k
\vert^2 \, ,
\ee 
where $\phi_k$ denotes the amplitude of the k'th Fourier mode. The above
expression is divergent both in the infrared and in the ultraviolet. The
ultraviolet divergence is the usual divergence of a free quantum field
theory and can be ``cured" by introducing an ultraviolet cutoff. In the
infrared, we will discard all modes with wavelength larger than the Hubble
radius at the beginning of inflation, since these modes are determined by
the pre-inflationary physics. We take these modes to contribute to the background.

At any time $t$ we can separate the integral in (\ref{tpf}) into the contribution of infrared and ultraviolet modes, the separation being defined by setting the physical wavelength equal to the Hubble radius. Thus, in an inflationary Universe the infrared phase space is continually increasing since comoving modes are stretched beyond the Hubble radius, while the ultraviolet
phase space is either constant (if the ultraviolet cutoff corresponds to a fixed physical wavelength), or decreasing (if the ultraviolet cutoff corresponds to fixed comoving wavelength). In either case, unless the spectrum of the initial fluctuations is extremely blue, two point functions such as (\ref{tpf}) will at later stages of an inflationary Universe be completely dominated by the infrared sector. In the following, we will therefore restrict our attention to this sector, i.e. to wavelengths larger than the Hubble radius.

In order to evaluate the two point functions which enter into the expressions for $\tau_{\mu \nu}$, we need to know the time evolution of the linear fluctuations $\phi_k$, which is given by the linear theory of cosmological perturbations \cite{MFB92}. On scales larger than the Hubble radius,
and for a time-independent equation of state, the well-known result for $\phi_k$ is
\be \label{phi} 
\phi_k(t) \, \simeq \, {\rm const} \, .
\ee
The Einstein constraint equations yield a relation between the metric $\phi$ and the matter fluctuations and $\delta \varphi $. 
\be 
\dot \phi +H\phi = 4\pi G\dot \varphi _0\,\delta \varphi \,.  \label{constr}
\ee
In most models of inflation, exponential expansion of the Universe results
because $\varphi _0$ is rolling slowly, i.e. 
\be 
\dot \varphi _0\simeq -{\frac{V^{\prime }}{3H}}\,,  \label{eom}
\ee
where a prime denotes the derivative with respect to the scalar matter field. Making use of (\ref{phi}), we can combine Eqs. (\ref{constr}) and (\ref{eom}) to obtain 
\be 
\delta \varphi =-{\frac{2V}{V^{\prime }}}\,\phi \,.  \label{constr2}
\ee
Hence, in the expressions (\ref{tzero}) and (\ref{tij}) for $\tau_{\mu \nu}$, all terms with space and time derivatives can be neglected, and we obtain
\be 
\rho _{br}\equiv \tau _0^0\cong \left( 2\,{\frac{{V^{\prime \prime }V^2}}{{%
V^{\prime }{}^2}}}-4V\right) <\phi ^2>  \label{tzerolong}
\ee
and 
\be 
p_{br}\equiv -\frac 13\tau _i^i\cong -\rho_{br} \,,  \label{tijlong}
\ee

The main result which emerges from this analysis is that the equation of state
of the dominant infrared contribution to the energy-momentum tensor $\tau_{\mu \nu}$ which describes back-reaction takes the form of a {\it negative cosmological constant} 
\be \label{result}
p_{br}=-\rho _{br} \,\,\, {\rm with} \,\,\, \rho_{br} < 0 \, .
\ee

The second crucial result is that the magnitude of $\rho_{br}$ increases as
a function of time. This is due firstly to the fact that, in an inflationary Universe, as time increases more and more wavelengths become longer than the Hubble radius and begin to contribute to $\rho_{br}$. A second reason for the growth of the absolute value of $\rho_{br}$ is that the amplitude of the individual modes $\phi_k$ changes as a consequence of the evolution and slow change in the equation of state of the background, as governed by the ``conservation law" \cite{zeta}  $\zeta = {\rm const}$ where
\be \label{cons}
\zeta \, = \, {2 \over 3}{{H^{-1} {\dot \phi} + \phi} \over {1 + w}} + \phi \, ,
\ee 
where $w = p/\rho$. In the models studied, the effect of the growth of the individual fluctuation modes is larger than the effect due to the increase in the phase space of infrared modes.
 
\section{Application to Chaotic Inflation}

To study the magnitude of back-reaction, we will
consider a single field chaotic inflation model \cite{Linde}   
with potential  
\be 
V(\varphi )={\frac 12}m^2\varphi ^2\,.  \label{pot}
\ee
Furthermore, we specify an initial state at a time $t_i$ in which the
homogeneous inflaton field has the value $\varphi _0(t_i)$ and  the
fluctuations are minimal. 

Using the values of $\phi_k$ on long wavelengths which are well known  \cite{MFB92}, the (dominant) infrared contribution to the correlator becomes
\be   \label{correl}
\langle \phi^2 (t) \rangle \, = \, \int_{k_i}^{k_t} \frac{dk}{k} |\phi_k|^2
\, = \, \frac{m^2 M_P^2}{32 \pi^4 \varphi_0^4 (t)} \int_{k_i}^{k_t} \frac{dk%
}{k} {\left[ \ln{\ \frac{H(t) a(t)}{k} } \right]}^2 \, ,
\ee
where $M_P$ is the Planck mass, $k_i$ corresponds to the physical infrared cutoff, and $k_t$ is the time-dependent comoving wavelength corresponding to the Hubble radius. Since the background evolution is known, the integral over $k$ can be done. The resulting back-reaction energy density $\rho_{br}$ can hence be calculated and compared with the background density $\rho_0$. The result is
\be   \label{result2}
{\frac{{\rho_{br}(t)} }{{\rho_0}}} \, \simeq \, {\frac{{3} }{{4 \pi}}} {\frac{{
m^2 \varphi_0^2(t_i)} }{{M_P^4}}} \left[ {\frac{{\varphi_0 (t_i)} }{{\varphi_0
(t)}}} \right]^4 \, .
\ee

As follows from (\ref{result2}),
back-reaction may lead to a shortening of the period of inflation. Without
back-reaction, inflation would end \cite{Linde} when $\varphi_0 (t) \, \sim \, M_P$. Inserting this value into (\ref{result2}), we see that if 
\be 
\varphi_0 (t_i) \, > \, \varphi_{br} \, \sim \, m^{-1/3} M_P^{4/3} \, ,
\ee
then back-reaction will become important before the end of inflation and may shorten the period of inflation. It is interesting to compare this value with the scale
\be 
\varphi_0 (t_i) \, \sim \, \varphi_{sr} \, = \, m^{-1/2} M_P^{3/2} \, ,
\ee
which emerges in the
scenario of stochastic chaotic inflation \cite{Starob,Slava}  as 
the ``self-reproduction" scale beyond which quantum fluctuations dominate
the evolution of $\varphi_0 (t)$. Notice that $\varphi_{sr} \gg \varphi_{br}$
since $m \ll M_P$. Hence, even in the case when self-reproduction does not take place, back-reaction effects can be very important.

\section{Speculations}

Since the back-reaction of cosmological fluctuations in an inflationary cosmology acts (see (\ref{result})) like a negative cosmological constant,
and since the magnitude of the back-reaction effect increases in time, one
may speculate \cite{RB98} that back-reaction will lead to a dynamical relaxation of the cosmological constant (see Tsamis \& Woodard \cite{TW} for similar considerations based on
the back-reaction of long wavelength gravitational waves).

The background metric $g_{\mu \nu}^{(0, br)}$ including back-reaction evolves as if the cosmological constant at time $t$ were
\be \label{effcosm}
\Lambda_{\rm eff}(t) \, = \, \Lambda_0 + 8 \pi G \rho_{br}(t)
\ee
and not the bare cosmological constant $\Lambda_0$. Hence we propose to identify (\ref{effcosm}) with a time dependent effective cosmological constant. Since $\vert \rho_{br}(t) \vert$ increases as $t$ grows, the effective cosmological constant will decay. Note that even if the initial magnitude  of the perturbations is small, eventually (if inflation lasts a sufficiently long time) the back-reaction effect will become large enough to cancel any bare cosmological constant.

Furthermore, we speculate that this dynamical relaxation mechanism for $\Lambda$ will be self-regulating. As long as $\Lambda_{\rm eff}(t) \, > \, 8 \pi G \rho_m(t)$, where $\rho_m(t)$ stands for the energy density in ordinary matter and radiation, the evolution of $g_{\mu \nu}^{(0, br)}$ is dominated by $\Lambda_{\rm eff}(t)$. Hence, the Universe will be undergoing accelerated expansion, more scales will be leaving the Hubble radius and the magnitude of the back-reaction term will increase. However, once $\Lambda_{\rm eff}(t)$ falls below $\rho_m(t)$, the background will start to decelerate, scales will enter the Hubble radius, and the number of modes contributing to the back-reaction will decrease, thus reducing the strength of back-reaction. Hence, it is likely that there will be a scaling solution to the effective equation of motion for $\Lambda_{\rm eff}(t)$ of the form
\be \label{scaling}
\Lambda_{\rm eff}(t) \, \sim \, 8 \pi G \rho_m(t) \,.
\ee
Such a scaling solution would correspond to a contribution to the relative closure density of $\Omega_{\Lambda} \sim 1$.
 
\section{Discussion}

We have summarized a new formalism to study the effect of linear cosmological perturbations on the cosmological background. This effect is expressed in terms of an effective energy-momentum tensor. We have shown that the back-reaction can be described in a way which is invariant under linear coordinate transformations. The issue of invariance under higher order transformations is crucial but still unresolved \cite{Unruh}. The most interesting result which emerges is that, in an inflationary background, the effective energy-momentum tensor which describes the back-reaction has the form of a negative cosmological constant. The absolute value of the induced effective energy density grows in time and, in a model with a long period of inflation, can become significant, which leads to the speculation that the effect may lead to a dynamical relaxation of the cosmological constant. However, the effective energy-momentum tensor defined in this work describes the effect of fluctuations on the homogeneous mode of the gravitational field. If the speculations in the previous section are to hold up, the analysis must be extended to give back-reaction effects on local quantities. Work on this issue is in progress.

\section*{Acknowledgments}

I thank my collaborators Raul Abramo and Slava Mukhanov for a very stimulating collaboration (they should not be blamed if the speculations of the final
section do not hold up!). I also thank A. Guth, W. Unruh and R. Woodard for many lively discussions. This work is supported in part by DOE Contract DE-FG0291ER40688, Task A, and by the US NSF collaborative research award NSF-INT-9312335.

\end{document}